\documentclass{PoS}

\usepackage{graphicx}
\usepackage{epsfig}
\usepackage{subfigure}

\def\rm #1{\mbox{\scriptsize #1}}
\def\eq#1{\begin{equation} #1 \end{equation}}
\def\eqarray#1{\begin{eqnarray} #1 \end{eqnarray}}
\newlength{\x}
\newlength{\y}
\newlength{\z}
\newcommand{\extralongrightarrow}[1]{{}_{\stackrel{\rule[0.45mm]{6mm}{0.09mm}\!\longrightarrow}{#1}}}

\title{Quark and Glue Momenta and Angular Momenta in the Proton --- a Lattice Calculation}

\ShortTitle{Quark and glue momenta and angular momenta}

\author{\speaker{{}K.F. Liu} $^{a}$, M. Deka $^{b,c}$, T. Doi$^d$, Y.B. Yang$^{e}$, B. Chakraborty$^a$, 
Y. Chen$^e$, S.J. Dong$^a$, \mbox{T. Draper$^a$}, M. Gong$^a$, H.W. Lin$^f$, D. Mankame$^a$, N. Mathur$^g$, and T. Streuer$^b$ \\
\llap{$^a$}Department of Physics and Astronomy, University of
Kentucky, Lexington, KY 40506 \\
\llap{$^{b}$} Institute for Theoretical Physics, University of Regensburg, 
  93040 Regensburg, Germany \\
\llap{$^{c}$} The Institute of Mathematical Sciences, Chennai 6000113, India \\
\llap{$^{d}$} Theoretical Research Division, Nishina Center, RIKEN, Wako 351-0198, Japan \\
\llap{$^{e}$} Institute of High Energy Physics, Chinese Academy of Sciences, 
Beijing 1000190, China \\
\llap{$^{f}$} Department of Physics, University of Washington, Seattle, WA 98195 \\
\llap{$^{g}$} Department of Theoretical Physics, Tata Institute of
  Fundamental Research, Mumbai 40005, India \\
E-mail: \email{liu@pa.uky.edu} 
\begin{center} ($\chi$QCD Collaboration) \end{center}}

\bigskip

\abstract{We report a complete calculation of the quark and glue momenta and
angular momenta in the proton.  These include the quark contributions from both
the connected and disconnected insertions. The calculation is carried out on a
$16^3 \times 24$ quenched lattice at $\beta = 6.0$ and for Wilson fermions with
$\kappa = 0.154, 0.155,$ and 0.1555 which correspond to pion masses at 650,
538, and 478 MeV. The quark loops are calculated with $Z_4$ noise and
signal-to-noise is improved further with unbiased subtractions. The glue
operator is comprised of gauge-field tensors constructed from the overlap
operator.  The $u$ and $d$ quark momentum/angular momentum fraction is
0.66(5)/0.72(5), the strange momentum/angular momentum fraction is
0.024(6)/0.023(7), and that of the glue is 0.31(6)/0.25(8). The orbital angular
momenta of the quarks are obtained from subtracting the angular momentum
component from its corresponding spin. As a result, the quark orbital angular
momentum constitutes 0.50(2) of the proton spin, with almost all it coming from
the disconnected insertion. The quark spin carries a fraction 0.25(12) and glue
carries a fraction 0.25(8) of the total proton spin.}

\FullConference{XXIX International Symposium on Lattice Field Theory \\
		 July 10-16 2011\\
		 Squaw Valley, Lake Tahoe, California}

\begin{document}

\section{Introduction}
Determining the contributions of quarks and gluons to the nucleon spin is one
of the challenging issues in QCD both experimentally and theoretically. Since
the contribution from the quark spin is small ($\sim$ 25\%) from deep inelastic
scattering experiments, it is expected that the rest should come from glue spin
and the orbital angular momenta of quarks and glue.

Lattice calculations of the quark orbital angular momenta have been carried out
for the connected insertions~\cite{Mathur:1999uf,Hagler:2003jd,bgh07,bee10} and
it was shown to be small~\cite{Mathur:1999uf} in the quenched calculation and
near zero in dynamical fermion calculations~\cite{Hagler:2003jd,bgh07,bee10}
due to the cancellation between those of the $u$ and $d$ quarks. Gluon helicity
distribution $\Delta G(x)/G(x)$ from both COMPASS and STAR experiments is found
to be close to zero~\cite{COMPASS-STAR-1011}. Furthermore, it is argued based
on analysis of single-spin asymmetry in unpolarized lepton scattering from a
transversely polarized nucleon that the glue orbital angular momentum is
absent~\cite{bg06}.  Thus, it appears that we have encountered a `Dark Spin'
scenario.

In this work, we give a complete calculation of the quark and glue momenta and
angular momenta.  The quark contributions in both the connected and
disconnected insertions are included. Combining with earlier work of the quark
spin, we obtain the quark orbital angular momenta. We find that indeed the $u$
and $d$ quark orbital angular momenta largely cancel in the connected
insertion. However, their contributions including the strange quark are large
(50\%) in the disconnected insertion due to the fact that the quark spin for
each of the $u$, $d$, and $s$ quarks in the disconnected insertion is large and
negative. We have been able to obtain the glue momentum and angular momentum
for the first time, mainly because the overlap operator is used for the
gauge-field tensor which is less noisy than that constructed from the gauge
links. The lattice renormalization of the quark and glue energy-momentum
operators are achieved through the momentum and angular momentum sum rules. Due
to these sum rules, the gravitomagnetic moment is zero as proven for composite
systems from the light-cone Fock representation~\cite{bhm01}.

\section{Formalism}

\subsection{Angular Momenta and Momenta for Quarks and Gluons}

The angular momentum operator in QCD can be expressed as a gauge-invariant
sum~\cite{ji97},
\eq{
\vec J_{\rm{QCD}} = \vec J_q + \vec J_g
  = \frac{1}{2} \vec\Sigma + \vec{L}_q + \vec{J}_g
\label{ang_op_def_split_1}
}
where $\vec J_q = \frac{1}{2} \vec\Sigma + \vec{L}_q$ and $\vec J_g$ are the
quark and gluon contributions respectively. $\vec\Sigma$ is the quark spin
operator, and $ \vec{L}_q$ is the quark orbital angular momentum operator.\
$\vec J_{q,g}$ can be expressed in terms of the energy-momentum tensor operator
through
\eq{
J_i^{q,g} = \frac{1}{2}\,\epsilon_{ijk}\,\int \, d^3x\, (\mathcal{T}_{4k}^{q,g}\, x^j
          - \mathcal{T}_{4j}^{q,g}\, x^k)
\label{ang_op_def_split_2}
} 
Similarly, the quark and glue momentum operators are
\eq{
P_i^{q,g} = \int \, d^3x\, \mathcal{T}_{4i}^{q,g},
\label{momentum_op}
}
where
\eq{
{\cal T}_{4i}^q =  \frac{-i}{4}\sum_f \overline {\psi}_f \lbrack \gamma_4
         \stackrel{\rightarrow}{D}_i + \gamma_i \stackrel{\rightarrow}{D}_4
        -  \gamma_4 \stackrel{\leftarrow}{D}_i 
        - \gamma_i \stackrel{\leftarrow}{D}_4 \rbrack \psi_f\ 
\label{q_contrib_def_1}
}
and
\eq{
{\cal T}_{4 i}^g = i \sum_{k=1}^3 G_{4k}\, G_{ki}
\label{g_contrib_def_1}
}
The matrix element of the energy-momentum tensor, $\mathcal{T}_{4i}$, between
two nucleon states with momenta, $p'$ and $p$, can be written as~\cite{ji97}
(in Euclidean space),
\eqarray{
(p,s | {\cal T}_{4i}^{q,g} | p',s')
  &=& \left(\frac{1}{2}\right) \bar{u}(p,s) \left[T_1(q^2)(\gamma_4\bar{p}_i 
   +  \gamma_i\bar{p}_4) 
   - \frac{1}{2m}T_2(q^2)(\bar{p}_4 \sigma_{i\alpha} q_{\alpha} 
   +  \bar{p}_i \sigma_{4\alpha} q_{\alpha})\right.\nonumber\\
  &-& \left.\frac{i}{m} T_3(q^2) q_4 q_i\right]_{q,g} u(p',s')
\label{mat_element_1}
}
where, $\bar{p} = \frac{1}{2}\, (p + p')$, $q_\mu = p_\mu - p'_\mu$, $m$ is the
mass of the nucleon, and $u(p,s)$ is the nucleon spinor.

By substituting Eq.~(\ref{mat_element_1}) into Eqs.~(\ref{ang_op_def_split_2})
and (\ref{momentum_op}) at $q^2 \rightarrow 0$ limit, one obtains
\eqarray{
J_{q,g} &=& \frac{1}{2} \left[T_1(0) + T_2(0)\right]_{q,g}, \label{ang_op_def_split_3}\\
\langle x\rangle_{q,g} &=& T_1(0)_{q,g}.
\label{momentum_fraction}
}

In this present study, there are two operators --- energy momentum tensors for
the quarks and glue. We can use the momentum and angular momentum sum rules to
perform the renormalization on the lattice
\eqarray{
&& Z_q(a) T_1(0)_q + Z_g(a) T_1(0)_g = 1, \label{mom_sum} \\
&& Z_q(a) \lbrack T_1(0)_q + T_2(0)_q\rbrack + Z_g(a) \lbrack T_1(0)_g 
+ T_2(0)_g\rbrack = 1. \label{ang_mom_sum}
}
After the lattice renormalization, one can perform perturbative mixing and
matching to the $\overline{\rm{MS}}$ scheme at $\mu = 2$ GeV.

It is interesting to note that from these equations, i.e.\ Eqs. (\ref{mom_sum})
and (\ref{ang_mom_sum}), one obtains the sum of the $T_2(0)$'s for the quarks
and glue to be zero, i.e.
\eq{
Z_q(a) T_2(0)_q + Z_g(a) T_2(0)_g = 0.
\label{T_2}
}
This shows that the total anomalous gravitomagnetic moment of the nucleon
vanishes. This has been proven by Brodsky {\it et al.}~\cite{bhm01} for
composite systems from the light-cone Fock representation and now it is shown
as a consequence of momentum and angular momentum conservation. We will see how
large the individual quark and glue contributions are.

Since $T_1(q^2)$ and $T_2(q^2)$ can have different $q^2$ behaviors, we shall
compute them separately at different $q^2$ values~\cite{Hagler:2003jd}, and
then separately extrapolate them to $q^2 \rightarrow 0$ for both the quark and
glue contributions.

\subsection{Glue Energy-Momentum Tensor Operator}  \label{field_tensor}

It is well-known that gauge operators from the link variables are quite
noisy. We adopt the glue energy-momentum tensor operator where the field
tensors are obtained from the overlap Dirac operator. The gauge-field tensor
has been derived from the massless overlap operator $D_{\rm
ov}$~\cite{Liu:2007hq}
\eq{
   T\!r_s  \left[\sigma_{\mu\nu} D_{\rm ov}(x,x)\right] =  c_T a^2G_{\mu\nu}(x) 
+ {\mathcal O}(a^3)
}
where $T\!r_s$ is the trace over spin. $c_T = 0.11157$ for $\kappa = 0.19$.
With this construction, we expect the ultraviolet fluctuations to be
suppressed, since $D_{\rm ov}$ is exponentially local and $D_{\rm ov}(x,x)$
involves gauge loops beginning and ending at $x$, which serves as smearing.

\section{Lattice Calculations and Numerical Parameters}
In order to obtain $J_{q,g}=\frac{1}{2} \left[T_1(0) + T_2(0)\right]_{q,g}$ and
$\langle x \rangle_{q,g} = T_1(0)_{q,g}$, we first calculate the three-point
functions, $G_{N {\cal T}_{4i} N}(\vec{p},t_2;\vec{q},t_1; \vec{p'},t_0)$ and
the two-point functions $G_{NN}(\vec{p},t_2)$. Here $\vec{p}$ is the sink
momentum, $\vec{q}$ is the momentum transfer, and $\vec{p}'$ is the source
momentum. $t_0$, $t_1$ and $t_2$ are the source, current insertion, and sink
time respectively. We then take the following ratios between three-point and
two-point functions, which involve the combinations of $T_1(q^2), T_2(q^2)$,
and $T_3(q^2)$
\eqarray{
  &&\frac{\mbox{Tr}\left[\Gamma_{e,m}  G_{N {\cal T}_{4i} N}
          (\vec{p},t_2;-\vec{q},t_1; \vec{p'},t_0)\right]}
         {\mbox{Tr}\left[\Gamma_e  G_{NN}(\vec{p},t_2)\right]} \times
    \sqrt{\frac{\mbox{Tr}\left[\Gamma_e  G_{NN}(\vec{p'},t_2-t_1+t_0)\right]}
               {\mbox{Tr}\left[\Gamma_e  G_{NN}(\vec{p},t_2-t_1+t_0)\right]}} \times \nonumber \\
  &&\sqrt{\frac{\mbox{Tr}\left[\Gamma_e  G_{NN}(\vec{p},t_1)\right]}
               {\mbox{Tr}\left[\Gamma_e  G_{NN}(\vec{p'},t_1)\right]}\cdot
    \frac{\mbox{Tr}\left[\Gamma_e  G_{NN}(\vec{p},t_2)\right]}
         {\mbox{Tr}\left[\Gamma_e  G_{NN}(\vec{p'},t_2)\right]}}\,
    \extralongrightarrow{t_1 \gg t_0, t_2 \gg t_1} \,\,\,
%   {}_{\stackrel{\rule[0.6mm]{6mm}{0.1mm}\longrightarrow}{t_1 \gg t_0,\, t_2 \gg t_1}} \,
    \frac{\left[a_1T_1(q^2) + a_2T_2(q^2) + a_3T_3(q^2)\right]}{4\sqrt{E_p(E_p+m)E_{p'}(E_{p'}+m)}} 
\label{ang_mom_1}
}
where $\Gamma_m = (-i)/2(1 + \gamma_4)\gamma_m\gamma_5$ is the spin polarized
projection operator, and $\Gamma_e = 1/2 (1 + \gamma_4)$ is the unpolarized
projection operator.  For the special case with $\vec{p} = 0$, one obtains the
combination of $\left[T_1 + T_2\right](q^2)$
\eq{
  \frac{\mbox{Tr}\left[\Gamma_m  G_{N {\cal T}_{4i} N}
        (\vec{0},t_2;-\vec{q},t_1; \vec{p'},t_0)\right]}
       {\mbox{Tr}\left[\Gamma_e  G_{NN}(\vec{0},t_2)\right]} \cdot
  \frac{\mbox{Tr}\left[\Gamma_e  G_{NN}(\vec{0},t_1)\right]}
       {\mbox{Tr}\left[\Gamma_e  G_{NN}(-\vec{q},t_1)\right]}\,\,\,
  \extralongrightarrow{t_1 \gg t_0, t_2 \gg t_1} \,\,\,
%   {}_{\stackrel{\rule[0.6mm]{6mm}{0.1mm}\longrightarrow}{t_1 \gg t_0,\, t_2 \gg t_1}} \,\,\, 
  \epsilon_{ijm} q_j \left[T_1 + T_2\right](q^2),
\label{ang_mom_2}
}
and the forward matrix element, which is obtained with the following ratio
\eq{
  \frac{\mbox{Tr}\left[\Gamma_e G_{N {\cal T}_{4i} N}
             (\vec{p},t_2;\vec{0},t_1; \vec{p},t_0)\right]}
       {|\vec{p}|\,\mbox{Tr}\left[\Gamma_e G_{NN}(\vec{p},t_2)\right]}\,\,\,
  \extralongrightarrow{t_1 \gg t_0, t_2 \gg t_1} \,\,\,
  \langle x \rangle.
\label{first_mom_rat_1}
}
Since the quark $T_1(q^2)$ and $T_2(q^2)$ for the connected insertion are shown
to have quite different $q^2$ behavior~\cite{Hagler:2003jd,bgh07,bee10}, we
will need to separately extrapolate $T_1(q^2)$ and $T_2(q^2)$ to $q^2
\longrightarrow 0$.  We combine several kinematics and with both the polarized
and unpolarized three-point functions to extract $T_1(q^2), T_2(q^2)$, and
$T_3(q^2)$ which appear as different combinations in the 3-point to 2-point
ratios in Eqs.~(\ref{ang_mom_1}) and then extrapolate $T_1(q^2)$ and $T_2(q^2)$
in $q^2$ to obtain $T_1(0)$ and $T_2(0)$. We can check this by comparing the
extracted $T_1(q^2) + T_2(q^2)$ against the $\left[T_1 + T_2\right](q^2)$
obtained directly from \mbox{Eqs. (\ref{ang_mom_2}) and (\ref{first_mom_rat_1})
at comparable $q^2$.}

The three-point functions for quarks have two topologically distinct
contributions in the path-integral diagrams --- one from connected (CI) and the
other from disconnected insertions
(DI)~\cite{Liu:1993cv,Liu:1998um,Liu:1999ak}. For DI, we sum over the current
insertion time, $t_1$, between the source and the sink time, i.e.\ from $t_1 =
t_0 +1$ to $t_2 -1$~\cite{Maiani:1987by,Dong:1995rx,Deka:2008xr,Doi:2009sq} in
order to increase the statistics. Similarly for the glue. Then the
corresponding ratios at large time separation for Eqs.~(\ref{ang_mom_1}),
(\ref{ang_mom_2}), and (\ref{first_mom_rat_1}) are $\frac{\left[a_1T_1(q^2) +
a_2T_2(q^2) + a_3T_3(q^2)\right]}{4\sqrt{E_p(E_p+m)E_{p'}(E_{p'}+m)}} \times
t_2 +\rm{const.}, \epsilon_{ijm} q_j \left[T_1 + T_2\right](q^2)_{q,g} \times
t_2 + \rm{const.}$, and $\langle x \rangle \times t_2 + \rm{const.}$
respectively.

We then extract the slopes and obtain $T_1(0), T_2(0)$, $\left[T_1 +
T_2\right](q^2)_{q,g}$ and $\langle x \rangle_{q,g}$ in the DI the same way as
was done for the CI.

We use $500$ gauge configurations on a $16^3 \times 24$ lattice generated with
Wilson action at $\beta=6.0$ in the quenched approximation. The values of the
hopping parameter we use are $\kappa = 0.154, 0.155$, and $0.1555$. The
critical hopping parameter, $\kappa_c = 0.1568$ is obtained by a linear
extrapolation to the zero pion mass~\cite{Dong:1995ec}. Using the nucleon mass
to set the lattice spacing at $a = 0.11$ fm, the corresponding pion masses are
$650(3), 538(4)$, and $478(4)$~MeV, and the nucleon masses are $1291(9),\
1159(11)$, and $1093(13)$~MeV, respectively. We use Dirichlet boundary
condition in the present work.\

In the case of DI, the quark loop is evaluated separately. We compute it
stochastically by using complex $Z_2$ noise vector~\cite{Dong:1993pk}. The
number of noise vectors we use is 500 on each gauge configuration. Also for the
case of quarks, we shall define two $\kappa$'s for the quark mass: $\kappa_v$
for valence quarks, and $\kappa_{\rm{sea}}$ for sea quarks. For the strange
quark current, we have fixed $\kappa_{\rm{sea}} = 0.154$ which is close to the
strange quark mass as determined from the $\phi$ meson mass, and $\kappa_v$
takes the values of $0.154, 0.155$, and $0.1555$. For up and down quarks, we
consider the cases with equal valence and sea quark masses, i.e.\
$\kappa_{\rm{sea}} = \kappa_v = 0.154, 0.155$, and $0.1555$.

In the case of glue, we construct the energy-momentum tensor from the gauge
field tensors which are calculated from the massless overlap operator, $D_{\rm
ov}$~\cite{Liu:2007hq,Doi:2008hp} as discussed in
Sec. \ref{field_tensor}. $D_{\rm ov} (x,x)$ at all space-time points are
estimated stochastically where we compute the color and spin indices exactly,
but we perform space-time dilution by a separation of two sites on top of
odd/even dilution. Therefore, the ``taxi-driver distance'' $= 4$ in our case.\
We use two $Z_4$ noise sources, and take the average of them on each
configuration.

Due to the stochastic estimation, there is noise in addition to that from the
gauge configurations. So, we adopt the following techniques to reduce the
error:
\begin{itemize}
\item
We use discrete symmetries to discard the unwanted part of the current and the
two-point functions when we correlate the current to the two-point functions to
construct the three-point functions~\cite{ber89,Draper:1988bp,Mathur:1999uf}.\
Combining $\gamma_5$-Hermiticity, parity and $CH$ transformation, we determine
whether the real or imaginary part of the current and the two-point functions
will contribute to the three-point functions.
\item
We employ unbiased subtraction to the noise estimation of the quark loop to
reduce the contributions from the off-diagonal matrix
elements~\cite{Thron:1997iy,Mathur:1999uf,Deka:2008xr}. We use four subtraction
terms ($\kappa D$, $\kappa^2 D^2$, $\kappa^3 D^3$ and $\kappa^4 D^4$). No
subtraction has been implemented for the estimation of the glue matrix
elements.
\item
We use multiple nucleon sources (in our case, 16) to increase the statistics.\
We correlate the nucleon propagators at different source locations with the
already computed quark loops which results in significant reduction of
errors~\cite{Deka:2008xr,Doi:2009sq}.
\end{itemize}

The error analysis is performed by using the jackknife procedure. The
correlation among different quantities are taken into account by constructing
the corresponding covariance matrices. In order to extract various physical
quantities, we use correlated least-$\chi^2$ fits.

\section{Numerical Results}
\begin{figure}[htbp]
\centering
\subfigure[]
{{\includegraphics[width=0.45\hsize]{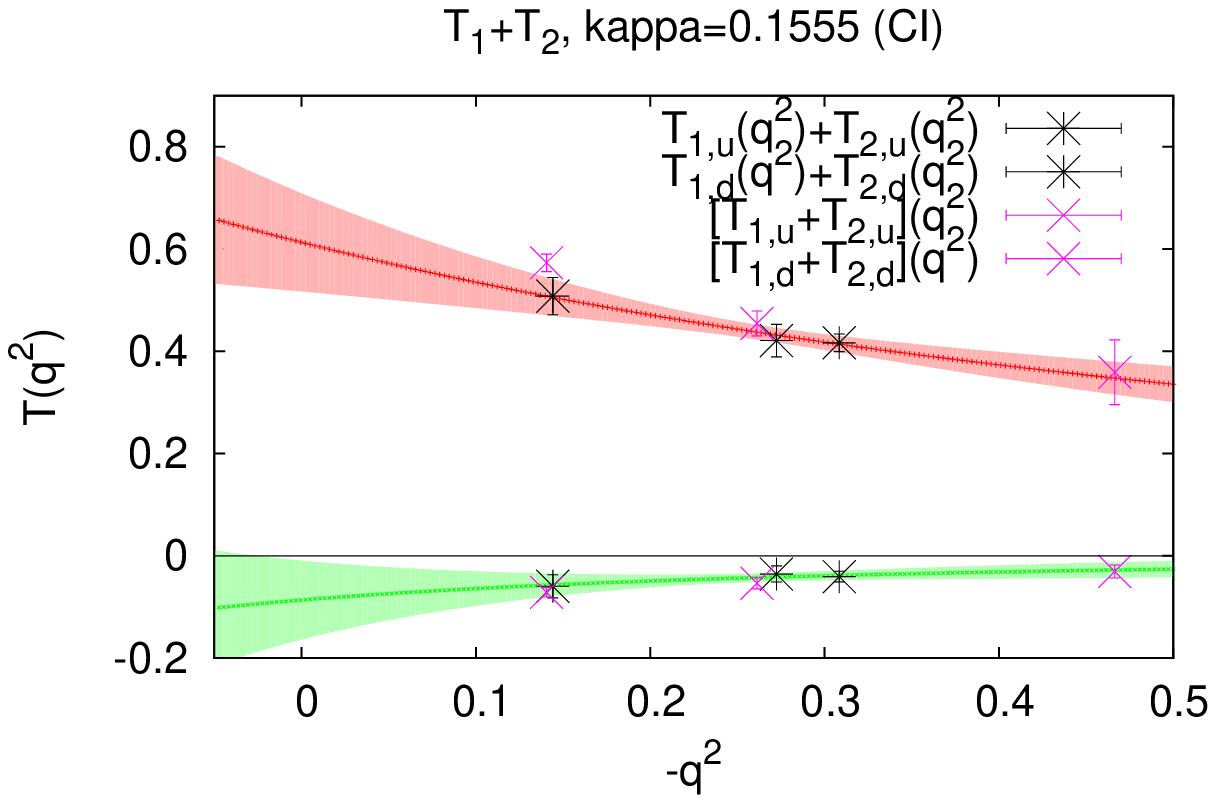}}
  \label{fig:quark_ang_mom_CT_1}}
\hspace{0.5cm}
\subfigure[]
{{\includegraphics[width=0.45\hsize]{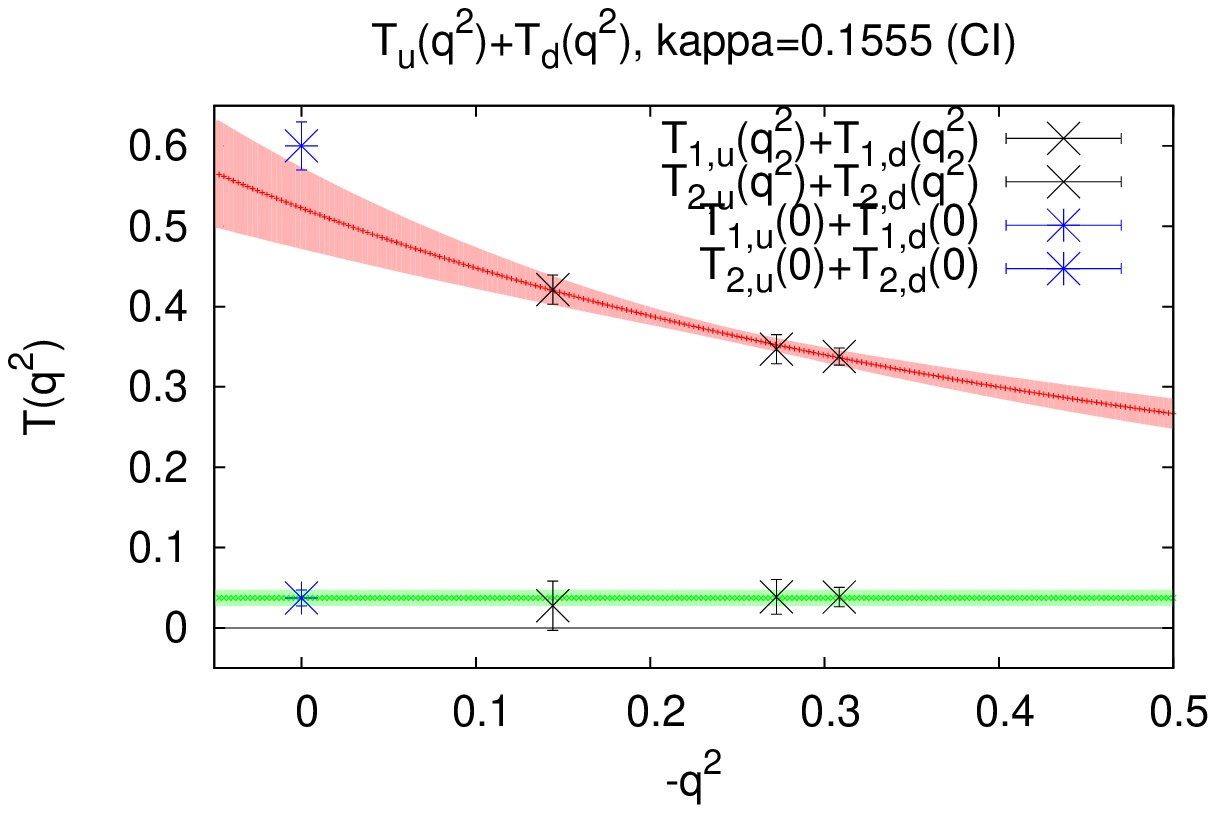}}
\label{fig:quark_ang_mom_CT_2}}
\caption{Results from $\kappa = 0.1555$: (a) The sum of $T_1(q^2)$ and
$T_2(q^2)$ extracted from Eq.~(\protect\ref{ang_mom_1}) with error bands from
the dipole fit is compared to $\left[T_1 + T_2\right](q^2)$ from
Eq.~(\protect\ref{ang_mom_2}) with slightly different $-q^2$ for $u$ and $d$
quarks in the CI.  (b) The sum of $u$ and $d$ quark contributions for
$T_1(q^2)$ and $T_2(q^2)$. The blue asterisk at $-q^2=0$ is $T_{1,u}(0) +
T_{1,d}(0)$ from Eq.~(\protect\ref{first_mom_rat_1}).  }
\end{figure}

We first present our results of the CI. In Fig.~\ref{fig:quark_ang_mom_CT_1},
we plot $T_{1,u}(q^2) + T_{2,u}(q^2)$ and $T_{1,d}(q^2) + T_{2,d}(q^2)$ as a
function of $-q^2$ for the case of $\kappa = 0.1555$, where $T_1(q^2)$ and
$T_2(q^2)$ are obtained from Eq.~(\ref{ang_mom_1}). We also plot $\left[T_{1,u}
+ T_{2,u}\right](q^2)$ and $\left[T_{1,d} + T_{2,d}\right](q^2)$ directly from
Eq.~(\ref{ang_mom_2}) at slightly different $-q^2$. We see that the latter
basically agrees with the error band of the former with a dipole fit in $-q^2$
within 2 sigmas. This is a cross check of our procedure of extracting
$T_1(q^2)$ and $T_2(q^2)$ from Eq.~(\ref{ang_mom_1}) which involves 3 to 4
equations of the combinations of $T_1(q^2), T_2(q^2)$, and $T_3(q^2)$ at
different $q^2$. We also show, in Fig.~\ref{fig:quark_ang_mom_CT_2},
$T_{1,u}(q^2) + T_{1,d}(q^2)$ and $T_{2,u}(q^2) + T_{2,d}(q^2)$ and their error
bands. Also plotted is $T_{1,u}(0) + T_{1,d}(0)$ from
Eq.~(\ref{first_mom_rat_1}). We see that its error is smaller than that from
the separately extrapolated $T_{1,u}(0)$ and $T_{1,d}(0)$. Thus we shall use
$T_{1,u}(0) + T_{1,d}(0)$ from Eq.~(\ref{first_mom_rat_1}) and combine with
$T_{2,u}(q^2) + T_{2,d}(q^2)$ to get the angular momentum for the CI.

\begin{figure}[htbp]
\centering
\subfigure[]
{{\includegraphics[width=0.5\hsize]{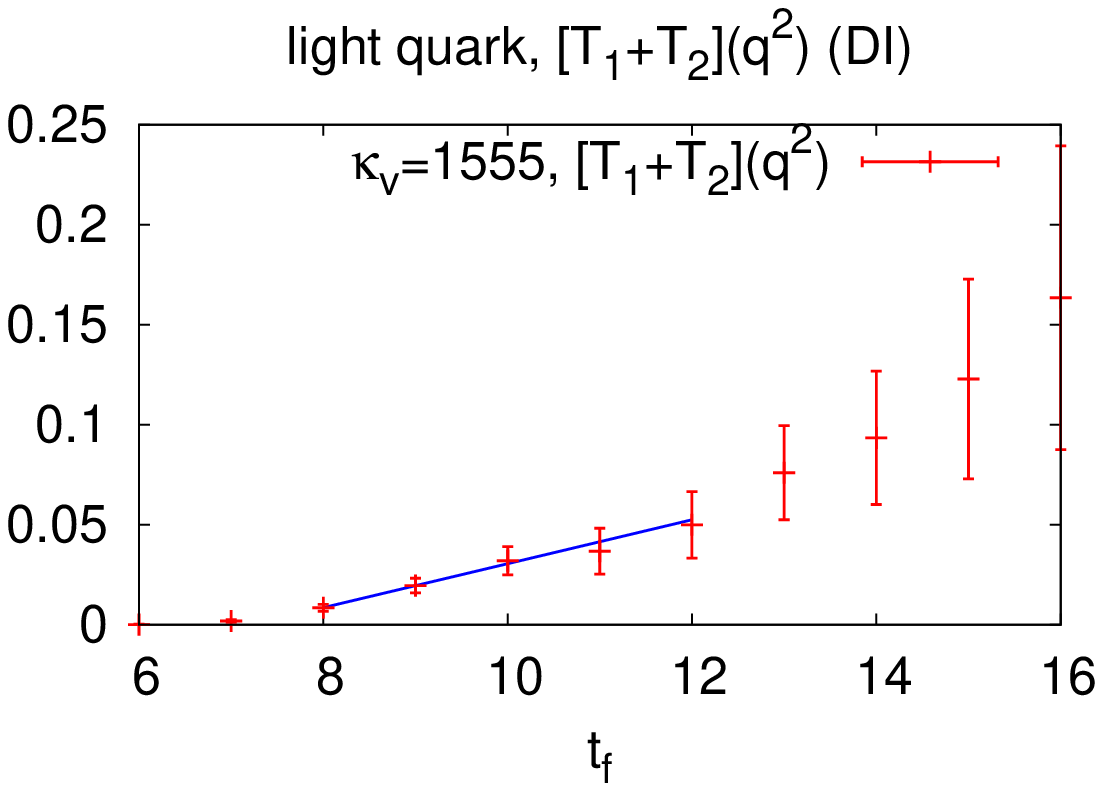}}
  \label{fig:slope_1555_DI}}
\subfigure[]
{{\includegraphics[width=0.5\hsize]{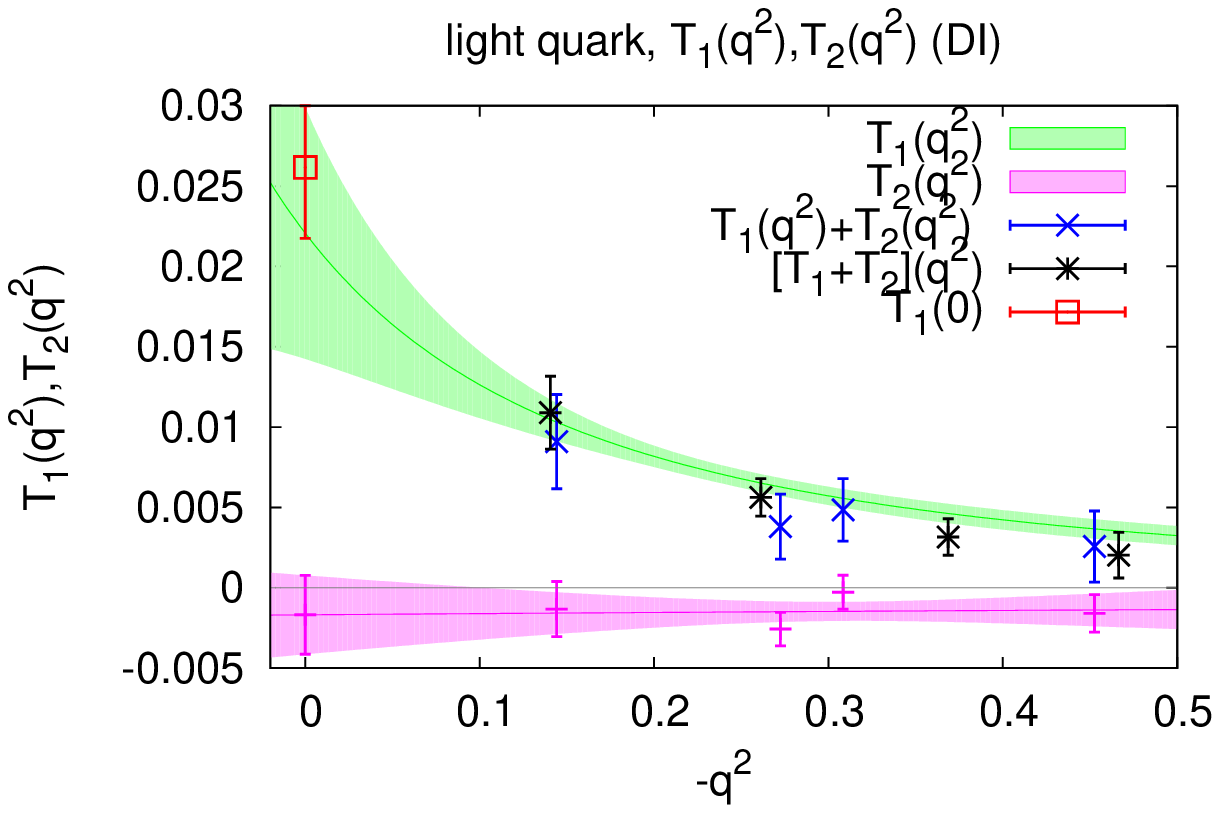}}
  \label{fig:T1T2_DI}}
%\vspace*{-0.6cm}
\subfigure[]
{{\includegraphics[width=0.5\hsize]{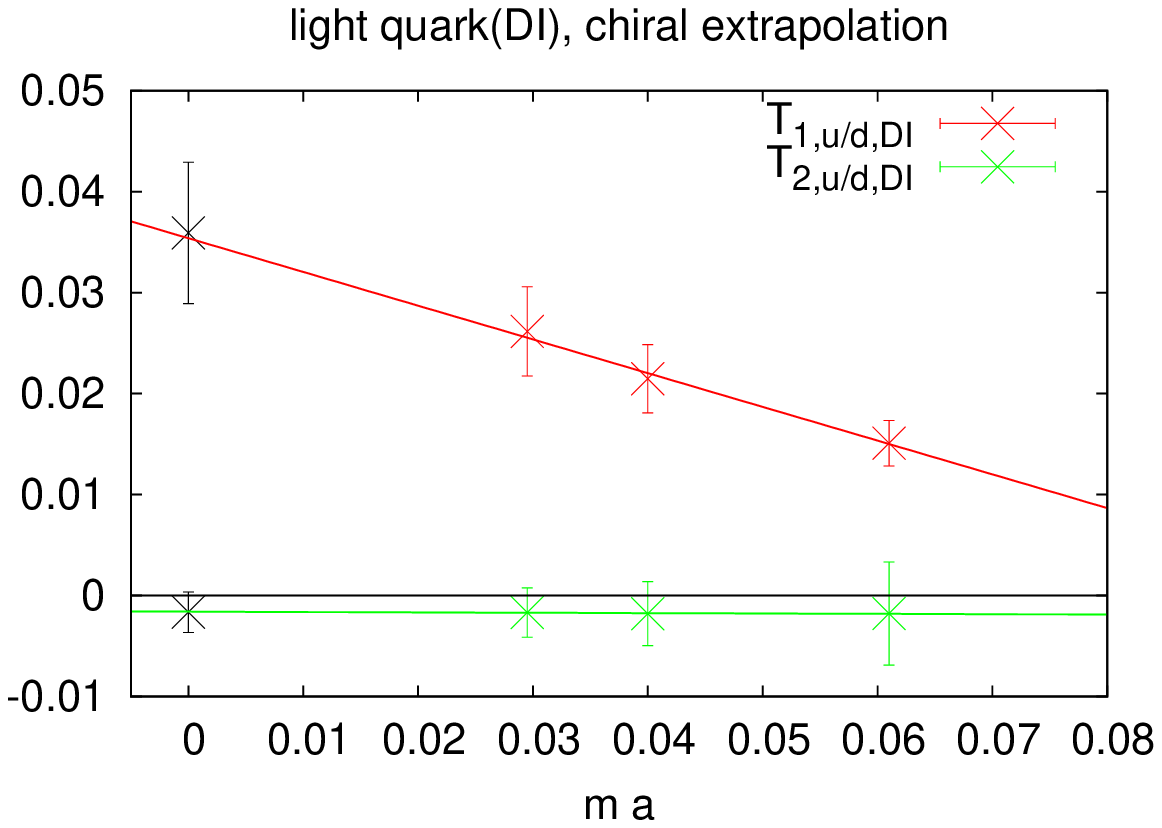}}
 \label{fig:chiral_fit_quark_ang_mom_DI}}
\caption{
  (a) The ratio of Eq.~(\protect\ref{ang_mom_2}) with $t_1$ summed between $t_0 +1$ 
and $t_2-1$ as a function of the sink time $t_2$. The slope is fitted to obtain
$\epsilon_{ijm} q_j \left[T_1 + T_2\right](q^2)_{q}$.
  (b) Separately extracted $T_1(q^2)$ and $T_2(q^2)$ are compared with 
      $\left[T_1+ T_2\right](q^2)$. $T_1(0)$ (red square) is from the forward matrix element. 
  (c) Chiral extrapolation of $T_1(0)$ and $T_2(0)$ for the $u/d$ quark. They are not renormalized.}
\label{fig:DI}
\end{figure}

For the DI with $\kappa_v = \kappa_{\rm{sea}} = 0.1555$ at $q^2a^2 = - 0.144$,
we show in Fig.~\ref{fig:slope_1555_DI} the ratio of Eq.~(\ref{ang_mom_2}) with
$t_1$ summed between $t_0 +1$ and $t_2-1$ and plotted against the sink time
$t_2$ so that the slope is $\epsilon_{ijm} q_j \left[T_1 + T_2\right](q^2)$. We
fit the slope from $t_2 =8$ where the ratio is dominated by the nucleon to $t_2
= 12$.  We plot $\left[T_1 + T_2\right](q^2)$ so obtained in
Fig.~\ref{fig:T1T2_DI} and compare them to $T_1(q^2) + T_2(q^2)$ extracted from
5 to 6 combinations of $a_1 T_1(q^2) + a_2 T_2(q^2) + a_3 T_3 (q^2)$.  We see
that they are consistent with each other within errors. The error bands are
from the dipole fits of $T_1(q^2)$ and $T_2(q^2)$. $T_1(0)$ (red square) is
from the forward matrix element which has smaller error than the $-q^2$
extrapolated $T_1(0)$. We shall combine it with the extrapolated $T_2(0)$ for
the angular momentum $J$ in the DI.  Finally, we perform a linear chiral
extrapolation to obtain $T_1(0) + T_2(0)$ for the $u/d$ quark at the chiral
limit. This is shown in Fig.~\ref{fig:chiral_fit_quark_ang_mom_DI}. For the
strange, we fix the quark loop at $\kappa_{\rm{sea}} = 0.154$ and extrapolate
$\kappa_v$ to the chiral limit.

\begin{figure}[htbp]
\centering
\subfigure[]
{{\includegraphics[width=0.5\hsize]{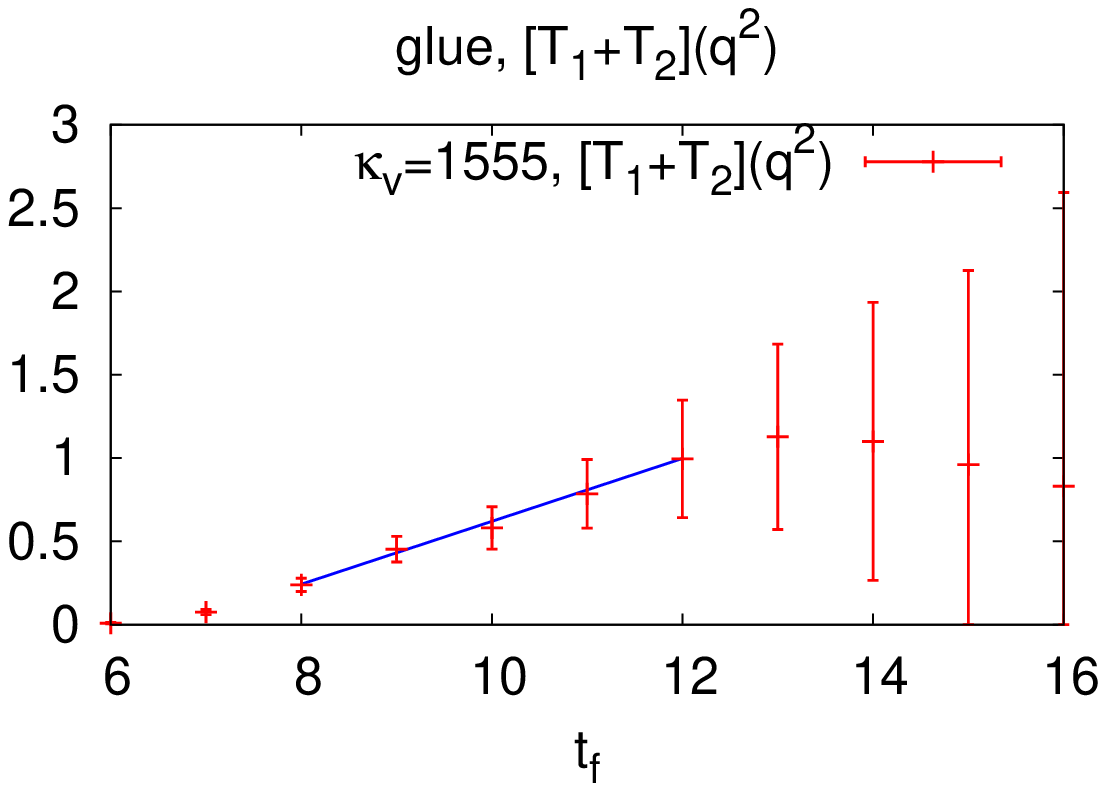}}
  \label{fig:slope_1555_g}}
\hspace{0.5cm}
\subfigure[]
{{\includegraphics[width=0.5\hsize]{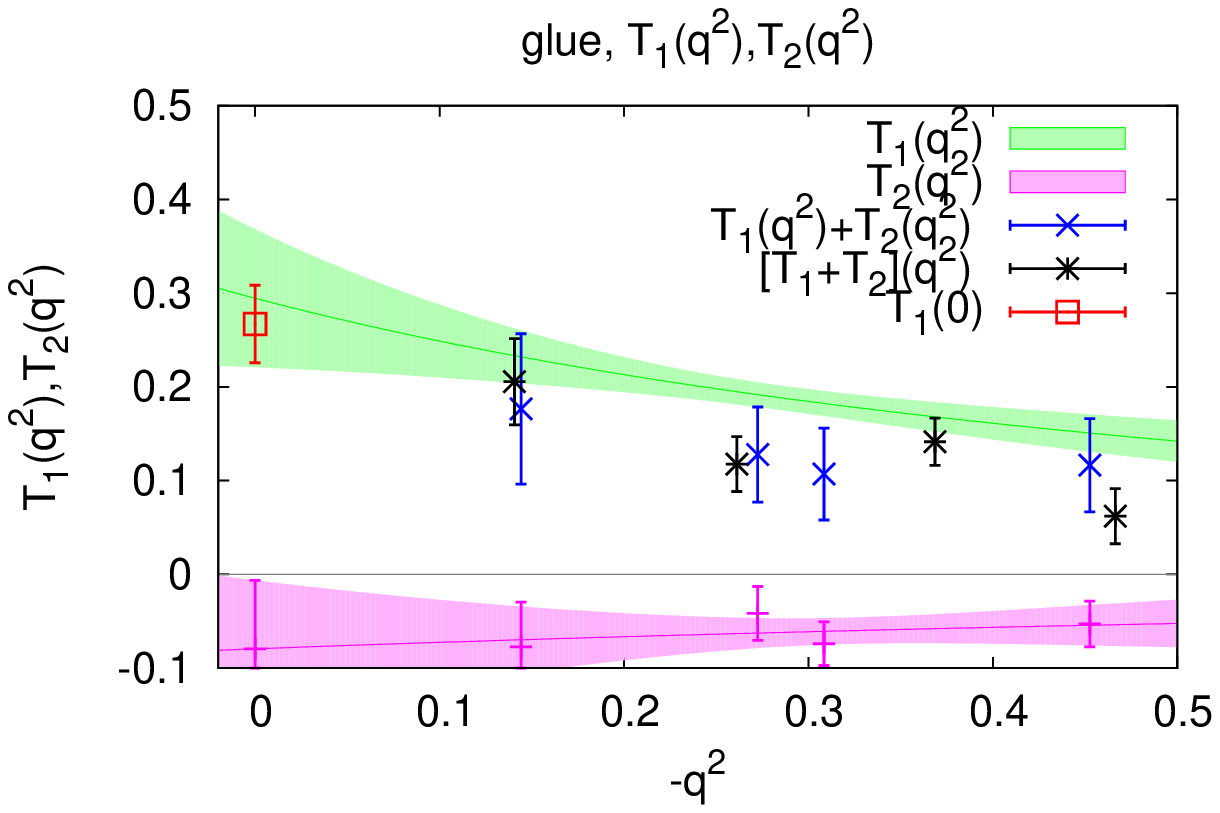}}
  \label{fig:T1T2_g}}
\vspace{1cm}
\subfigure[]
{{\includegraphics[width=0.5\hsize]{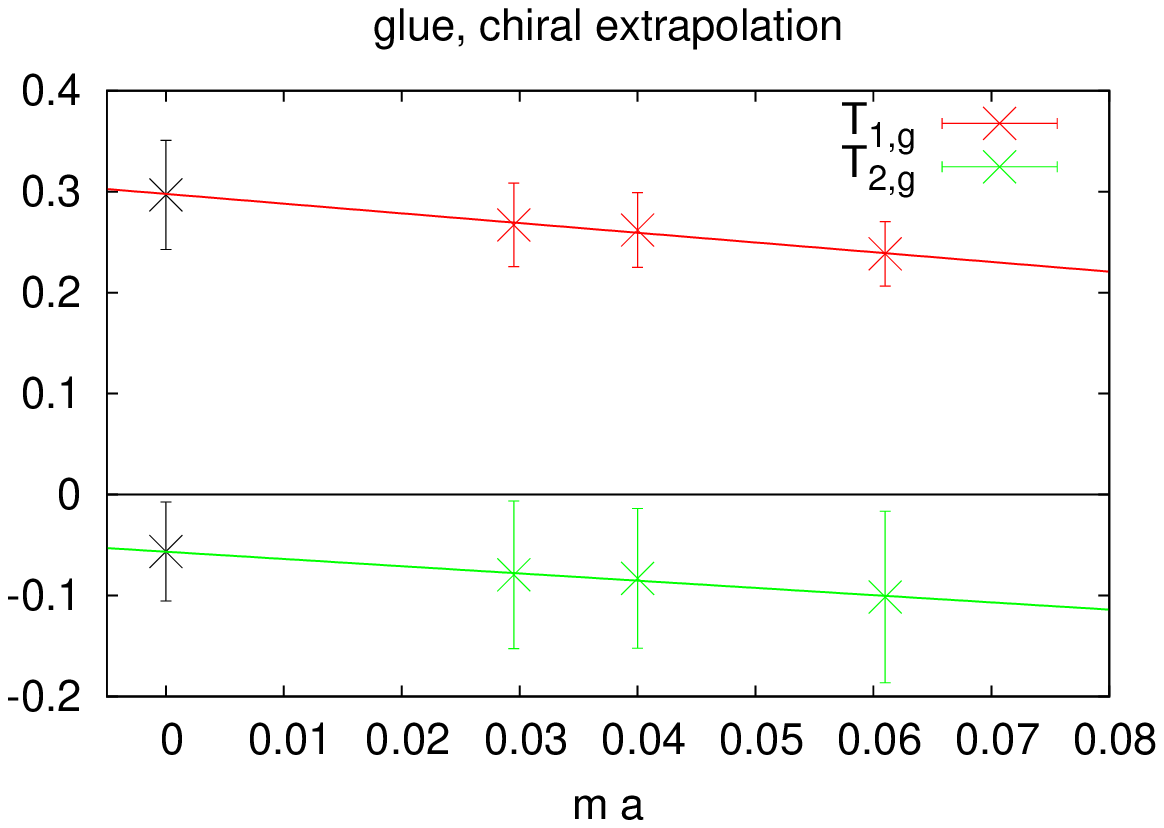}}
 \label{fig:chiral_fit_glue_ang_mom}}
\caption{
 The same as in Fig.~\protect\ref{fig:DI} for the glue.}
\label{fig:glue}
\end{figure}
%\vspace{2cm}
%

We perform similar analysis for the glue momentum and angular momentum.  They
are plotted in Figs.~\ref{fig:slope_1555_g},~\ref{fig:T1T2_g}
and~\ref{fig:chiral_fit_glue_ang_mom}.

With all the quark and glue momenta and angular momenta, we carry out lattice
renormalization through the sum rules in Eqs.~(\ref{mom_sum}) and
(\ref{ang_mom_sum}). We obtain $Z_q(a) = 1.05$ and $Z_g(a) = 1.05$. This shows
that both the lattice operators, particularly the glue energy-momentum tensor
from the overlap operator, are `natural' and close to the continuum. In
Table~\ref{tab:chiral}, we list the quark momentum fractions $\langle x\rangle
= T_1(0)$ in the CI ($u$ and $d$) and the DI ($u/d$ and $s$) and that of the
glue. We also list the corresponding $T_2(0)$ and angular momentum $2J = T_1(0)
+ T_2(0)$.

 \begin{table}[htb]
 \caption{Lattice renormalized values with renormalization constants $Z_q$=$Z_g$=1.05.} \label{tab:chiral}
 \begin{center}
 \begin{tabular}{|c||cc|cccc|}
\hline
                    & CI($u$)       & CI($d$)      & CI($u+d$) & DI($u/d$)   & DI($s$)     & Glue         \\
\hline
$\langle x \rangle$ &    0.428(40)  &    0.156(20) & 0.586(45) &    0.038(7) &    0.024(6) &    0.313(56) \\
$T_2(0)$            &    0.297(112) & $-$0.228(80) & 0.064(22) & $-$0.002(2) & $-$0.001(3) & $-$0.059(52) \\
$2J$                &    0.726(118) & $-$0.072(82) & 0.651(50) &    0.036(7) &    0.023(7) &    0.254(76) \\
$g_A$               &    0.91(11)   & $-$0.30(12)  & 0.61(8)   & $-$0.12(1)  & $-$0.12(1)  &  ---         \\
$2L$                & $-$0.18(16)   &    0.23(15)  & 0.04(9)   &    0.16(1)  &    0.14(1)  &  ---         \\
\hline
 \end{tabular}
 \end{center}
 \end{table}

We see from Table~\ref{tab:chiral} that the strange momentum fraction $\langle
x\rangle_s = 0.024(6)$ is in the range of uncertainty of $\langle x\rangle_s$
from the CTEQ fitting of the parton distribution function from experiments
which is $0.018 < \langle x\rangle_s < 0.040$~\cite{Lai:2007np}. The glue
momentum fraction of 0.313(56) is smaller than, say, the CTEQ4M fit of 0.42 at
$Q = 1.6$ GeV~\cite{Lai:1997CTEQ}. We expect the glue momentum fraction to be
larger than the present result when dynamical configurations with light
fermions are used in the calculation. From Figs.~\ref{fig:quark_ang_mom_CT_2}
and \ref{fig:T1T2_g} and Table~\ref{tab:chiral}, we find that the central
values of $T_2(0)$ for $u/d$ and $s$ in the DI are small and consistent with
zero. $T_{2,u}(0) + T_{2,d}(0)$ in CI is positive, while $T_{2,g}(0)$ is
negative. With the renormalization constants $Z_q(q)$ and $Z_g(a)$ fitted to be
both positive (and near unity), they cancel giving a null anomalous
gravitomagnetic moment. The fact that their unrenormalized magnitudes are
almost identical is consistent with the finding that $Z_q(q)$ and $Z_g(a)$ are
the same in size within errors.

In analogy to $F_2(0)$ which is known as the anomalous magnetic moment of the
nucleon, $T_2(0)$ is termed the anomalous gravitomagnetic moment and has been
shown by Brodsky {\it et al.} to vanish for composite systems~\cite{bhm01}. We
have now verified that this is the consequence of the momentum and angular
momentum sum rules.

The flavor-singlet $g_A^0$, which is the quark spin contribution to the
nucleon, has been calculated before on the same lattice~\cite{Dong:1995rx}. We
can subtract it from the total angular momentum fraction $2J$ to obtain the
orbital angular momentum fraction $2L$ for the quarks.  As we see in
Table~\ref{tab:chiral}, the orbital angular momentum fractions $2L$ for the $u$
and $d$ quarks in the CI have different signs and they add up close to zero,
0.04(9). This is the same pattern seen with dynamical fermions configurations
with light quarks~\cite{Hagler:2003jd,bgh07,bee10}. The large $2L$ for the
$u/d$ and $s$ quarks in the DI is due to the fact that $g_A^0$ in the DI is
large and negative, i.e.\ $-$0.12(1) for each of the three flavors. All
together, the quark orbital angular momentum constitutes a fraction 0.50(2) of
the proton spin. The majority of it comes from the DI. The quark spin fraction
of the nucleon spin is 0.25(12) and glue angular momentum contributes a
fraction 0.25(8).

\section{Summary}

In summary, we have carried out a complete calculation of the quark and glue
momentum and angular momentum in the nucleon on a quenched $16^3 \times 24$
lattice with three quark masses. The calculation includes both the connected
insertion (CI) and disconnected insertion (DI) of the three-point functions for
the quark energy-momentum tensor. We used complex $Z_2$ (or $Z_4$) noise to
estimate the quark loops in the DI and the gauge-field tensor from the overlap
operator in the glue energy-momentum tensor. We find that we can obtain
reasonable signal for the glue operator constructed from the overlap Dirac
operator. After chiral extrapolation, we use the momentum and angular momentum
sum rules to determine the lattice renormalization constants which turn out to
have a `natural' size close to unity, $Z_q(a)=1.05$ and $Z_g(a)=1.05$. The
lattice renormalized momentum fractions for the quarks are 0.586(45) for the CI
and 0.100(12) for the DI. The glue momentum fraction is 0.313(56). We have
shown that the anomalous gravitomagnetic moment Eq.~(\ref{T_2}) vanishes due to
the momentum and angular momentum conservation.

After subtracting from the angular momentum $2J$ the quark spin ($g_A^0$) from
a previous calculation on the same lattice~\cite{Dong:1995rx}, we obtain the
orbital angular fraction $2L$. In the CI, we find that the $u$ quark
contribution is negative, while the $d$ quark contribution is positive. The sum
is small, 0.04(9). This behavior is about the same as observed in dynamical
calculation with light quarks~\cite{Hagler:2003jd,bgh07,bee10}. The majority of
the quark orbital angular momentum comes from the DI, because the quark spin
from the DI is large and negative for each of the three flavors. In the end, we
find the quark orbital angular momentum, the quark spin, and glue angular
momentum fractions of the nucleon spin are 0.50(2), 0.25(12), and 0.25(8)
respectively.

We are in the process of calculating the perturbative matching to the
$\overline{\rm{MS}}$ scheme with mixing so that we can quote our results in
$\overline{\rm{MS}}$ scheme at 2 GeV.

\bigskip

\noindent {\large\bf Acknowledgment} 

\bigskip

This work is partially support by U.S. DOE Grant No. DE-FG05-84ER40154 and the
Center for Computational Sciences of the University of Kentucky.  The work of
M. Deka is partially supported by the Institute of Mathematical Sciences, India.

%\centerline{\bf  REFERENCES} %\vskip -15pt

\end{document}